# Angular Dependence of Vortex Annihilation Fields in Asymmetric Co Dots

Randy K. Dumas,[1] Thomas Gredig,[2,3] Chang-Peng Li,[2,§] Ivan K. Schuller[2] and Kai Liu[1,*]

[1]*Physics Department, University of California, Davis, CA 95616*
[2]*Physics Department, University of California - San Diego, La Jolla, CA 92093*
[3]*Department of Physics and Astronomy, California State University, Long Beach, CA 90840*

## Abstract

Shape asymmetries in nominally circular nanomagnets provide a potential means for vortex chirality control. However, in realistic arrays their effects are challenging to probe since asymmetric magnetization reversal processes are often averaged to include distributions over all angles. Here we investigate how shape asymmetry influences the vortex reversal in arrays of sub-micron edge-cut Co dots. We find that the vortices can be manipulated to annihilate at particular sites under different field orientations and cycle sequences. The vortex annihilation field and degree of chirality control depend sensitively on the angular position of the applied field relative to the flat edge of the dots. For small angles, the major loop annihilation field is significantly larger than that found from the half loop and the vortex chirality can be well controlled. At intermediate angles the chirality control is lost and an interesting crossover in the annihilation field is found: the half loop actually extrudes outside of the major loop, exhibiting a larger vortex annihilation field. At large angles the annihilation fields along major and half loops become degenerate.

**I.     Introduction**

Control over domain structures in magnetic nanoelements is critical to the understanding and applications of such materials.[1-4] In particular, magnetic vortices in sub-micron patterns have gained considerable interest in recent years due to their unique reversal mechanisms and potential applications in ultrahigh density patterned magnetic recording media.[5-10] The vortices are characterized by an in-plane magnetization with clockwise or counter-clockwise *chirality* and a central core with out-of-plane magnetization (up or down *polarity*). The ability to control the different vortex configurations within a single nanomagnet introduces alternative data storage possibilities.[11, 12] Typically in circular dots the vortex annihilation field is independent of where the vortex core is expelled from the dot. In realistic assemblies of dots, variations in dot shape, size, and intrinsic anisotropy inevitably exist and affect the reversal processes.[13] In particular, dot asymmetry has been shown to lift the degeneracy in vortex chirality, therefore providing a means for *chirality control*.[14-16] Recently, in studies of arrays of 67 nm Fe nanodots we have found distinct annihilation fields depending on which side of the dot annihilation occurred.[17, 18] The primary cause is the asymmetry in the dot shape due to deviations from perfect circles. In these Fe nanodots, as is typical in nanomagnet arrays, the slight shape asymmetry is randomly distributed, leading to asymmetric reversal in individual dots but overall isotropic behavior in the array. A key question is how the asymmetry influences the vortex reversal when its orientation is varied relative to the applied field. For example, how does the vortex nucleation/annihilation field change and is the chirality control always maintained? In this work we explore these issues in arrays of asymmetric Co dots where the circular shape in all the dots has been broken in the same fashion in order to gain insight into the reversal processes.

## II. Experiments

Polycrystalline arrays of Co dots were fabricated on naturally oxidized Si substrates with standard electron beam lithography and lift off techniques, in conjunction with magnetron sputtering. A scanning electron microscopy (SEM) image is shown in Fig. 1(a). The edge-cut dots are nominally 40 nm in thickness and 685 nm in diameter. They form a square array over a 100×100 µm² area, with a center-to-center separation of 900 nm. The pattern used to create each dot during e-beam writing is a regular dodecagon with three of the sides removed [Fig. 1(a) inset], thus creating an asymmetry. Photo-emission electron microscopy (PEEM) studies were carried out at beamline 11.0.1 of the Advanced Light Source (ALS). For comparison, arrays of circular Co dots are also examined, 40 nm in thickness and 870 nm in diameter, with a center-to-center-spacing of 1 µm. Remanent state images of both types of dots are shown in Figs. 1(b) and 1(c), after saturating the dots to the right. All the dots reverse the magnetization via a single vortex. Most of the edge-cut dots exhibit a counter-clockwise chirality [Fig. 1(b) inset],[19] while the circular dots show a random distribution of chirality.

Magnetic hysteresis loops were measured at room temperature using the magneto-optical Kerr effect (MOKE) on a Durham Magneto Optics NanoMOKE2 magnetometer.[5] The beam was focused to about a 30 µm diameter spot size, capturing the average reversal behavior of ~$10^3$ dots. The direction of an in-plane applied magnetic field was varied relative to the flat edge of the dots, which is defined as 0°. At each angle, major loops were measured between ±650 Oe and half loops were measured over 650 Oe – 0 Oe – 650 Oe, both with a field spacing of 2-4 Oe. At a field sweep rate of 11 Hz, typically ~$10^3$ loops were averaged to obtain a single hysteresis curve. Following prior procedures,[20] the first-order reversal curve (FORC) method was employed for selected angles. After positively saturating the sample the applied field was reduced to a given reversal field $H_R$, the magnetization $M$ was then measured back to positive saturation thereby tracing out a FORC. This process was repeated for more negative reversal fields until negative saturation is reached. A mixed second order derivative of the magnetization $M$ ($H$, $H_R$) was used to generate the FORC distribution $\rho \equiv -\partial^2 M(H,H_R)/2\partial H \partial H_R$, which capture the irreversible vortex nucleation/annihilation events.[21]

The experimental results were also compared with micromagnetic simulations performed with the OOMMF code.[22, 23] Material parameters suitable for these polycrystalline Co dots were used (saturation magnetization $M_S =1.4\times10^6$ A/m and exchange stiffness $A = 3\times10^{-11}$ J/m).[24] A weak residual uniaxial anisotropy of $K_1=1.1\times10^4$ J/m$^3$ with an easy axis along the flat edge of the dot was found experimentally and included in the simulations. A SEM image of an actual dot was used to construct the simulated dot, as shown in Fig. 4(c), thus reproducing the rounded edges from the lithography process.[25]

### III. Results and Discussions

The magnetic hysteresis loops of the Co dots are similar at different angles - all have the characteristic pinched shape with zero remanence and abrupt magnetization jumps that correspond to the vortex nucleation and annihilation.[5] Representative sets of major and half loops are shown in Fig. 2 for three angles. For clarity, only the first quadrant is shown [one full loop is shown as the outer boundary of FORC's in Fig. 3(a)]. At 0º, the vortex annihilation along the half loop, marked by the abrupt magnetization jump to positive saturation, occurs much sooner than that along the major loop [Fig. 2(a)]. Surprisingly, at 60º [Fig. 2(b)] the opposite behavior is observed where the annihilation along the half loop occurs noticeably later than that along the major loop. This leads to an unusual behavior where the half loop, a particular minor loop, extrudes outside of the major loop. This unusual pattern is a direct consequence of the dot asymmetry and the changing energy landscape during field cycling. The vortex annihilation along the half loop faces a higher energy barrier than along the full loop. Finally, at 90° [Fig. 2(c)] no discernable difference in vortex annihilation is observed. This trend was also qualitatively reproduced in simulated major and half loops, Fig. 2 (d-f), for the same representative angles. A comparison of the calculated micromagnetic energies reveals that the demagnetization and Zeeman energies play a dominant role during vortex nucleation and annihilation, while the weak anisotropy is the least significant.

A related asymmetry-driven vortex annihilation field was previously inferred in arrays of 67 nm Fe nanodots.[17, 18] Vortex annihilation fields along first order reversal curves also showed a crossover as vortices were annihilated from opposite sides of the dots. Consequently, a pronounced negative-positive-negative trio of features were found at the lower right corner of the FORC distribution $\rho(H, H_R)$,[17] where a positive peak in $\rho$ was accompanied by two adjacent negative valleys, highlighting the effect of shape asymmetry. However, the exact angular dependence of the annihilation field could not be resolved due to the random distribution of the shape asymmetry in the Fe nanodots. For comparison, we have carried out FORC analysis on the Co edge-cut dots discussed here.

At 0° the FORC distribution is characteristic of reversal via a vortex state.[17, 18] As shown in Fig. 3(b), a first prominent peak highlighted by region 1 at $(H, H_R) \sim$ (430 Oe, 220 Oe) corresponds to the annihilation of vortices from the flat side of the dot and essentially maps out the irreversible processes along the half loop, whose ascending branch is simply a FORC with $H_R=0$. A second peak highlighted by region 2 at (-220 Oe, -450 Oe) corresponds to the nucleation of vortices from negative saturation. Highlighted in region 3 at ~ (400 Oe, - 450 Oe) is a pronounced negative-positive pair of features, unlike the aforementioned trio of features observed in Fe nanodots.[17, 18] Note that along successive FORC's with more and more negative reversal fields, the vortex annihilation field approaching positive saturation moves progressively higher; over the applied field range of 350 Oe – 500 Oe, the slope of the FORC's first decreases and then increases, leading to respectively negative and positive values of $\rho$. The behavior has also been previously observed in simulated FORC's on edge-cut Fe dots and is due to the difference in annihilation fields,[17] which for instance can be easily observed between major and half loops. This interesting annihilation behavior becomes clear in the FORC and half loop analysis but is hidden when analyzing the major hysteresis loops alone. At 60º, a family of FORC's is shown in Fig. 3(a). The corresponding FORC distribution [Fig. 3(c)] resembles that at 0º, except for a weak negative-positive-negative trio of features in region 3 [Fig. 3(c) inset]. This set of features, which is almost identical to that found in the Fe dots discussed earlier and is caused by some of the FORC's extruding outside of the major loop.[26] The weak intensity is a manifestation of the small differences in annihilation fields along successive FORC's. At 90º, the two main peaks remain in the FORC distribution [Fig. 3(d)]; however, the feature in region 3 has faded away since once vortices have nucleated the annihilation field along subsequent FORC's remains the same (the trace amount of a residual feature is due to the small variations in the array). In the previously studied Fe dots, where all shape asymmetries were randomly distributed and averaged over,[17, 18] the three reversal behaviors typified by the FORC diagrams in Figs. 3(b-d) would all contribute to the observed negative-positive-negative trio of features. These results also demonstrate that shape asymmetry has a distinct effect on the vortex annihilation field, depending on the field cycle sequence, and can be turned on and off by varying the angular positions.

The annihilation field along major and half loops can be determined quantitatively from the field at which the magnetization jumps abruptly, i.e., where the M-H curve has a maximum slope. The angular dependence of the annihilation field extracted from the derivative of the measured loops is shown in Fig. 4(a). Three distinct regions, represented by the vortex annihilation behavior shown in Fig. 2, are found: at low angles the annihilation field along the half loop is significantly smaller than that along the major loop; at intermediate angles, especially over 55º-65º, the half loop annihilation field is slightly larger; at even higher angles, approaching 90º, the two annihilation fields converge. The resultant angular dependence[27] has a crossover region, roughly corresponding to the angular positions when the applied field passes through the corners of the flat edges of the dots.

Micromagnetic simulations also show the three distinct regions in the angular dependence, as shown in Fig. 4(b), qualitatively reproducing the measured results. Since the vortex nucleation along both major and half loops is the same, the difference in annihilation field is better illustrated by examining the vortex core annihilation sites found from simulations, as shown in Fig. 4(c). At small simulated angles (<30º), the vortex annihilation site is well defined due to chirality control achieved with an asymmetric dot [Fig. 1(b)].[14, 15] During reversal from saturation the magnetization preferentially buckles towards the flat edge of the dot, assisted by the demagnetizing field. A vortex is nucleated from the flat edge of the dot, and subsequently annihilated from the rounded edge of the dot along a major loop. However, if the field sweep is stopped at zero field and reversed towards positive saturation (i.e. tracing out a half loop) the vortex core must annihilate from the flat edge of the dot. For angles larger than 30º, simulations qualitatively reproduce the crossover region where the half loop annihilation field is larger than the major loop value. We find that the chirality control is lost and the annihilation site is either at a corner of the flat edge or the more rounded edge of the dot. For angles near 45º, annihilation from the corner (major loop) occurs at a smaller field than the rounded edge of the dot (half loop), due primarily to exchange energy gains, leading to the crossover behavior. Approaching 90º the dot asymmetry no longer plays a significant role as the vortex core moves parallel to the flat edge and the reversal mimics that of a symmetric dot. The annihilation sites along major and half loops become degenerate. Note that although these simulations are illustrative of the typical behavior in a single dot, some differences in the details from experimental results are still expected since the actual arrays of dots do have finite variations in their characteristics.[19]

It is worthwhile to examine the dipolar interactions in the array, which could potentially lead to a magnetostatically induced anisotropy. Novosad *et al.*[8] studied square array of permalloy dots and found that as the interdot distance decreased the nucleation and annihilation fields shifted towards zero field and the magnetostatic interactions between dots began to play an important role. They also showed that the interactions reduced as the dot aspect ratio (dot thickness $L$/ radius $R$) decreased. Guslienko calculated the fourfold anisotropy constant in rectangular arrays of dots as a function of normalized interdot distance.[7] For our dots, the normalized interdot spacing $\delta=d/R=0.63$ and the aspect ratio is $L/R=0.11$, where $d$ is the edge-to-edge dot spacing. According to Refs. 7 and 8, our dots are approaching the non-interacting regime and the magnetostatically induced anisotropy due to the array layout is not appreciable. In our OOMMF simulations of 3×3 and 5×5 arrays of dots, the reversal behaviors are nearly identical, and are qualitatively similar to those found in a single dot, as shown in Fig. 5. Only a very small reduction in the nucleation/annihilation fields is observed in the array simulations. Experimentally, only in an array of symmetric circular Co dots with a much smaller normalized interdot spacing of $\delta=0.27$, do we find evidence of the magnetostatically induced anisotropy.

Therefore we conclude that the present asymmetric dot arrays with $\delta=0.63$ are largely non-interacting.

## IV.    Conclusions

In summary, we have found that in asymmetric Co dots the vortex annihilation field and degree of chirality control depend sensitively on the angular position of the applied field relative to the flat edge of the dots. For small angles, the vortex is more easily expelled from the flat edge of the dots along a half loop than from the rounded edge of the dots along a major loop. The large difference between annihilation fields can be used to identify the vortex chirality. At intermediate angles the chirality control is lost and the opposite trend is observed. Along the half loop vortex annihilation from the rounded edge of the dots is harder than that from the dot corners along the major loop. Finally, at large angles approaching 90º the dot asymmetry is effectively removed as the vortex core moves parallel to the flat edge. Our results demonstrate an intrinsic effect of the shape asymmetry and illustrate how the vortices can be manipulated to annihilate at particular sites under certain field orientations and cycle sequences.

**Acknowledgements**

This work has been supported by CITRIS, the Alfred P. Sloan Foundation, and AFOSR. K.L. acknowledges a UCD Chancellor's Fellowship. The ALS is supported by the Director, Office of Science, Office of Basic Energy Sciences, of the U.S. Department of Energy under Contract No. DEAC02-05CH11231. We thank Andreas Scholl, Andrew Doran, Yayoi Takamura, Erika Jiménez, and Lifang Shi for technical assistance.

[27] The upward trend in the annihilation field is due to the weak residual uniaxial anisotropy found in the samples.

**Figure Captions**

Fig. 1. (Color online)  (a) SEM image of arrays of Co dots with a horizontal flat edge.  The inset shows the pattern used by the e-beam writer to create each dot and the 0º orientation of the sample.  PEEM images of a typical portion of (b) the edge-cut dots and (c) reference circular dots at zero field after saturating the dots to the right.  All dots are in the single-vortex state. The chirality is controlled in (b), as shown in the inset, and random in (c).

Fig. 2. (Color online)  (Left) Measured major and half loops using MOKE, and (Right) simulated major and half loops with the applied field at (a, d) 0º, (b, e) 60º and (c, f) 90º relative to the flat edge of the dots.

Fig. 3. (Color online)  (a) Measured FORC's and (b-d) FORC distributions with the applied field at (b) 0º, (a,c) 60º and (d) 90º relative to the flat edge of the dots.  Circles in (b) highlight the 3 regions of FORC features. Inset in (c) shows a zoom-in view of the negative-positive-negative set of features.

Fig. 4. (Color online) Angular dependence of the annihilation fields from both major (solid squares) and half loops (open circles) extracted from (a) MOKE measurements and (b) micromagnetic simulations.  Error bars are included to indicate the field spacing of the measurements and the lines are guides to the eye.  The locations of the simulated vortex core annihilation sites are shown in (c).  Lines are used to highlight the annihilation sites for each angular position.

Fig. 5. (Color Online) Simulated hysteresis loops for a 5×5 array (blue triangles), 3×3 array (red open circles), and single dot (black squares).

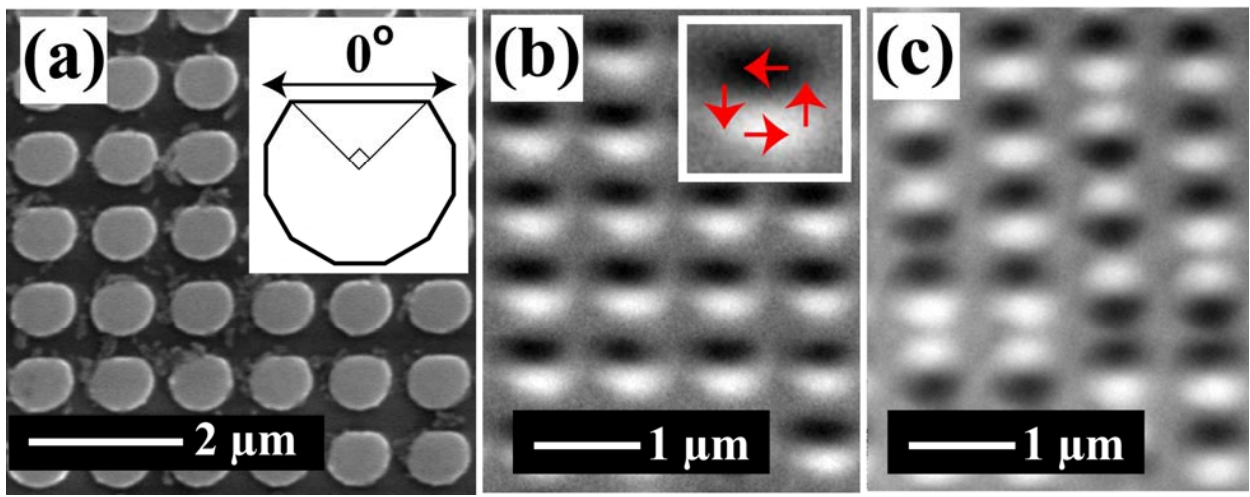

**Fig. 1. Dumas et al.**

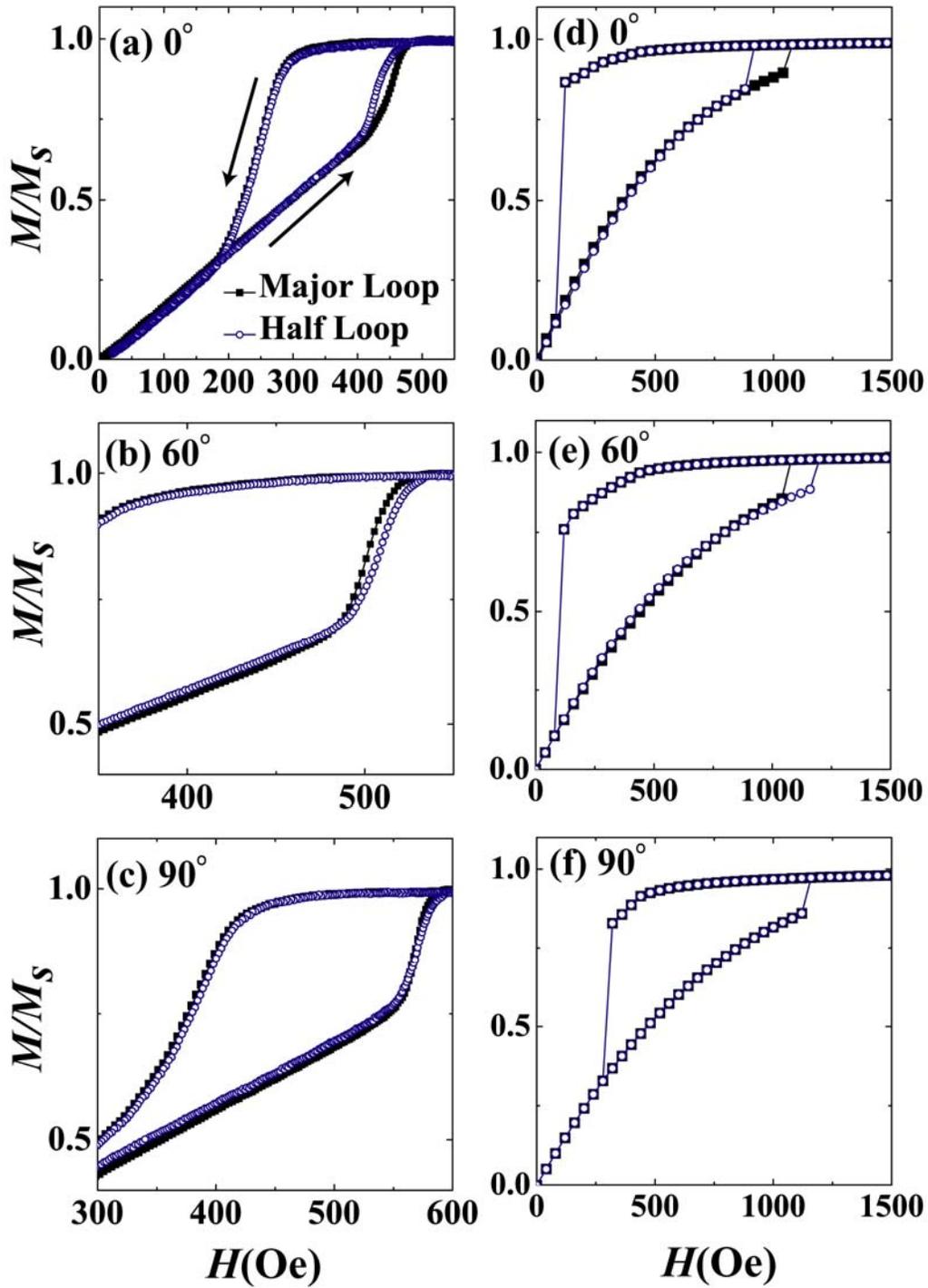

**Fig. 2. Dumas et al.**

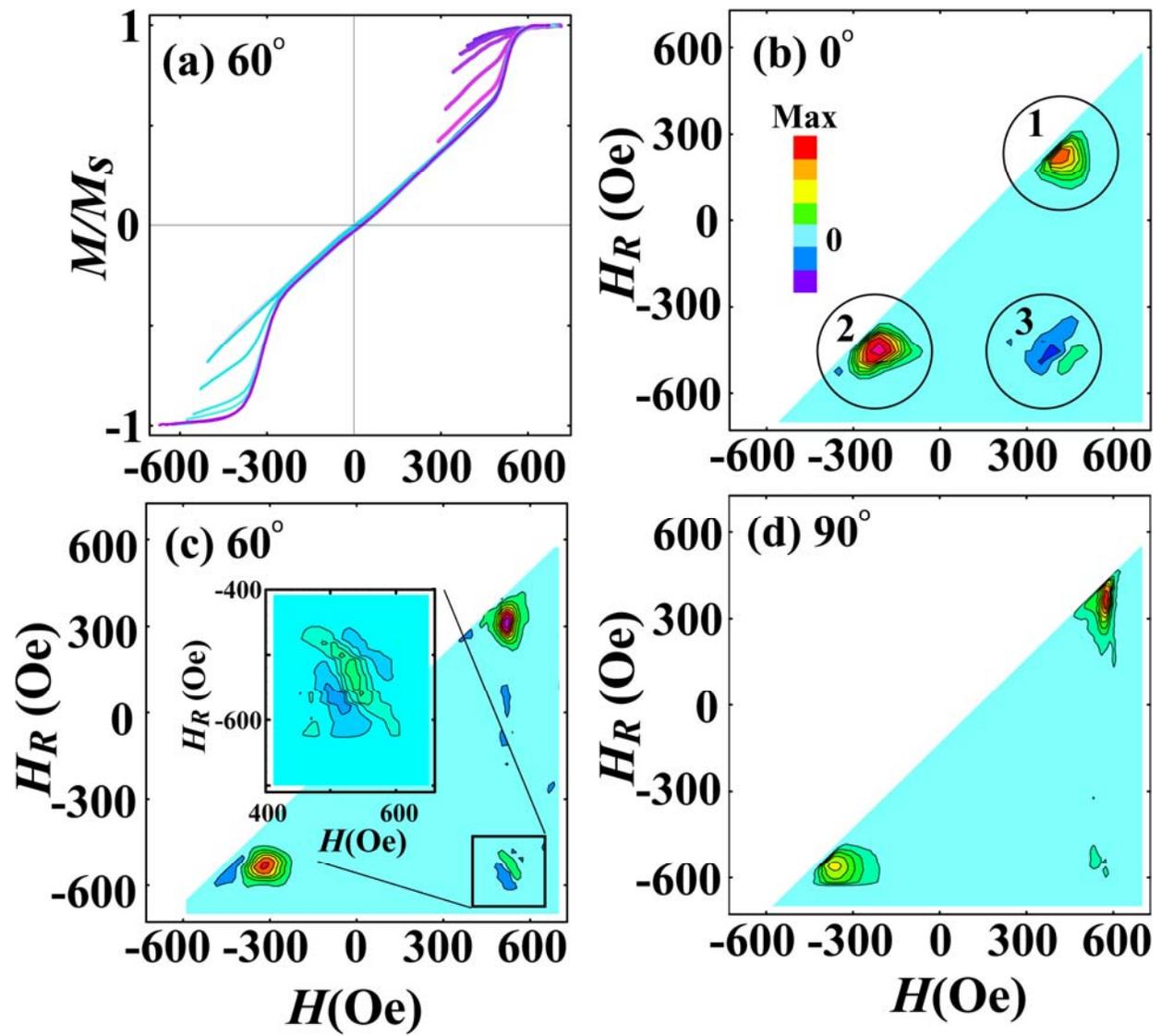

**Fig. 3. Dumas et al.**

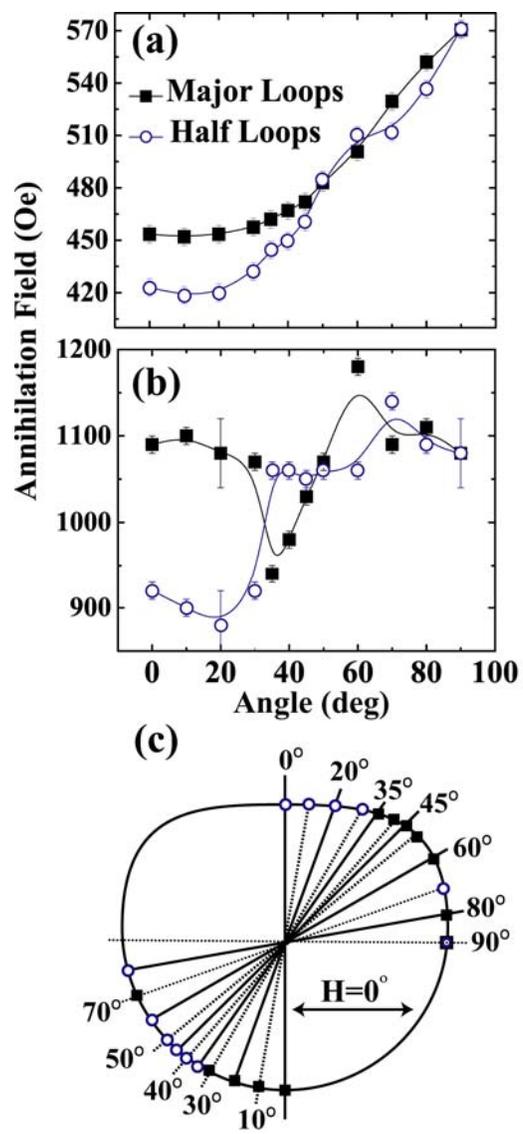

**Fig. 4. Dumas et al.**

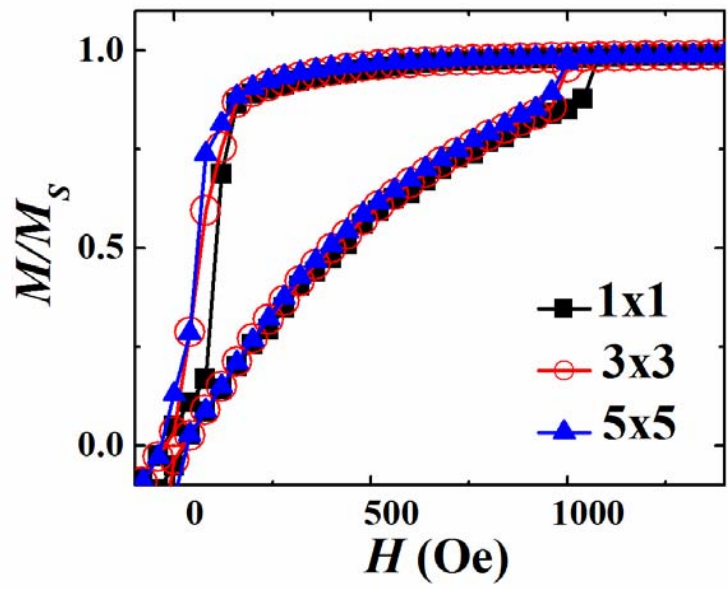

**Fig. 5. Dumas et al.**